\documentclass[12pt,oneside,a4paper]{book} 

\usepackage{natbib}
\usepackage{times,amsmath,float}
\usepackage{amssymb}
\usepackage{graphicx}
\usepackage{subfigure}
\usepackage{amsfonts}
\usepackage{epsfig}
\usepackage{anysize}
\usepackage{xtab}
\usepackage{lscape}
\usepackage{geometry}
\usepackage{longtable}
\usepackage{appendix}
\usepackage{setspace}
\usepackage{epstopdf}
\usepackage{chngcntr}
\usepackage{booktabs}
\usepackage{multirow}
\usepackage{caption,booktabs,array}
\usepackage{float}
\usepackage{adjustbox}
\usepackage{url}
\usepackage{multicol}
\usepackage{placeins}
\usepackage{pdfpages}
\usepackage{multicol}
\usepackage{titling}

\usepackage{xcolor}
\usepackage{graphicx}

\usepackage{titlesec}
\titleformat{\chapter}[display]{\normalfont\normalsize\centering\upshape}{\MakeUppercase{\chaptertitlename}\ \thechapter}{8pt}{\normalsize}
 
  \titleformat{\section}
  {\normalfont\normalsize\centering\upshape}{}{1em}{}

  \titleformat{\subsection}
{\normalfont\normalsize\centering\upshape}{}{1em}{}

\titlespacing{\chapter}{0pt}{-50pt}{0.3cm}

\usepackage{fancyhdr}
\usepackage[justification=justified,singlelinecheck=false]{caption}
\usepackage[labelsep=period]{caption}

\marginsize{4cm}{2.5cm}{2.5cm}{2.5cm}

\counterwithout{figure}{chapter}
\counterwithout{table}{chapter}
\counterwithout{footnote}{chapter}
\counterwithout{equation}{chapter}


\begin{document}

 \setlength{\parindent}{2.1em}
\doublespacing 

\frontmatter

\begin{center}

\vspace{7cm}
\LARGE Inequality in Turkey: Looking Beyond Growth
\vspace{5cm}

\Large Ipek Nazire Ergul\\
\&\\
\Large Bayram Cakir
\vspace{5cm}

\Large October 2019
\end{center}

\pagebreak 

\chapter{ABSTRACT}
\begin{center}
Inequality in Turkey: Looking Beyond Growth\\
\end{center}
\textbf{ } \\

\noindent 
This paper investigates the relationships between economic growth, investment in human capital and income equality in Turkey. The conclusion drawn based on the data from the OECD and the World Bank suggests that economic growth can improve income equality depending on the expenditures undertaken by the government. As opposed to the standard view that economic growth and income inequality are positively related, the findings of this paper suggest that other factors such as education and healthcare spending are also driving factors of income inequality in Turkey. The proven positive impact of investment in education and health care on income equality could aid policymakers who aim to achieve fairer income equality and economic growth, in investment decisions.

\pagebreak
\singlespacing

\chapter{TABLE OF CONTENTS}

\vspace*{0.4cm}
\doublespacing
\noindent CHAPTER 1: INTRODUCTION \dotfill 1 \hspace{1cm}\\
\noindent CHAPTER 2: LITERATURE REVIEW \dotfill 5 \hspace{1cm}\\
\noindent CHAPTER 3: THE DATA \dotfill 8 \hspace{1cm}\\
\noindent CHAPTER 4: GDP GROWTH and INCOME INEQUALITY \dotfill 10 \hspace{1cm}\\
\noindent CHAPTER 5: EDUCATION SPENDING and INCOME EQUALITY\dotfill 10 \hspace{1cm}\\
\noindent CHAPTER 6: HEALTH CARE SPENDING and INCOME EQUALITY\dotfill 10 \hspace{1cm}\\
\noindent CHAPTER 7: CONCLUSION \dotfill 25 \hspace{1cm}\\
\noindent REFERENCES \dotfill 27 \hspace{1cm}\\

\mainmatter
\chapter{INTRODUCTION}
\vspace*{0.4 cm}
\noindent Unfortunately, inequality has affected every society throughout history. While some countries experienced less severe income inequalities, some such as Ethiopia and Rwanda experienced much worse. In Ethiopia, more than 20\%  of the population lives below the poverty line and a significant number of the population cannot access clean water, electricity, health care, and education. Some even cannot afford basic needs such as adequate food, shelter, and clothing.  Exclusion from these resources causes inefficiencies in production and allocation of resources. Uneducated and unhealthy individuals cannot contribute to economic growth. Moreover, income inequality has social and political implications. High rates of crime, stress, suicide, political instability, and corruption all result from and feed economic inequality which results in a never endless cycle. This is known as the poverty cycle. The poverty cycle occurs when low incomes result in limited savings, because individuals spend their income on basic consumption needs, which then results in low investments in physical, human and natural capital, and therefore low productivity which in turn gives rise to low growth in income. The poverty cycle is transmitted from generation to generation which makes it nearly impossible for countries to break out of the cycle and achieve income equality on their own. \

Pure income equality suggests that each quintile in a country should possess the 20 \% of the real GDP generated in the economy. This is visualized in the Lorenz curve as the perfect income equality line. In most countries, there is income inequality which suggests that wealth is concentrated in the hands of a small percentage of the population. While a small percentage of the population lives in prosperity, the bottom half of the population lives in poverty. However, it is almost impossible to achieve complete pure income equality. Therefore, the focus is distributing income on the basis of equity rather than equality. Most countries restore income equality by direct provision of goods and services, subsidies, progressive taxation, and transfer payments. These methods aim to redistribute income by taking from high-income earners and giving it to low-income earners in the forms mentioned above. Some factors can help the distribution of income in a country and the role of economic growth in this process has been highly debated by economists. \

The relationship between inequality and growth has been debated since the late 20th century and it is still a widely discussed topic. However, there is still no consensus on whether there is a trade off between inequality and growth. While some economists suggest that economic growth increases income inequality, some argue the opposite. However, there are a significant number of economists that agree with both viewpoints. This paper argues that growth, still being a factor, is not the main driving factor that affects income inequality in a country. And only focusing on economic growth, growth of physical capital, won't lead to a more equal distribution of income in a society. For economic growth to bring economic development there must be investments made in human capital.  Education and health care are two distinct factors that affect the quality of human capital, therefore, the inequality in a country. Increases in the quantity and quality of education and health care increase the efficiency, productivity, and employability of workers. When workers are healthier and more educated, they produce more in less time and employers are more inclined to request more labor. Therefore, economic growth should be supported with increased investments in human capital including merit goods.\

This paper focuses on the relationship between economic growth and inequality and concludes that economic growth is not the main factor leading to income inequality in Turkey. It first analyzes the relationship between the Gini Coefficient and the GDP growth. It concludes that even though the data suggests that there is a positive relationship between the Gini Coefficient and the GDP growth it is not enough to say that growth hinders equality. This paper highlights the importance of education and health care on income inequality. And emphasizes that if GDP growth occurs through investments in human capital income inequality can decrease due to the reasons mentioned above. Moreover, the consequences that arise when there is a lack of government provision of these goods are explored. Lastly, the paper concludes that various factors affect income equality in Turkey. This paper can aid policymakers in deciding which capital to invest in. By demonstrating the importance of public investments in education and healthcare, the results could encourage policymakers to invest in human capital rather than physical capital to achieve economic development and income equality.\

The following section of this paper provides a literature review on the relationship between economic growth and income inequality. It put forwards various claims from different economists. It goes into depth on papers that analyze this relationship in relation to Turkey. Then, an overview of the data used in the paper is described. The implications and references to its source are explained in the section. The main body of the paper analyzes the relationship between the Gini coefficient and per capita growth, the consequences of education and health on economic development and inequality. In conclusion, the paper reaches the conclusion that income equality in Turkey is affected by many factors including but not limited to primary school enrollment, literacy rate, spending on education and spending on health care.\

\clearpage

\chapter{LITERATURE REVIEW}
\vspace*{0.4 cm}
\doublespacing
\noindent The question of whether poor benefit from economic growth has been explored by many economists. While there is a conventional argument that revolves around the idea that there is a trade-off between rapid growth and reduced inequality, there is some research that points out the positive correlation between rapid growth and inequality. While most research is focused on the United States or other European countries, there is limited research on the relationship between growth and inequality in Turkey.\

The economists that support the view that an unequal distribution of income is necessary for, or the likely consequence of, rapid economic growth base their beliefs on two fundamental explanations. Kaldor (1978) supports this view by stating that a high level of savings is imperative for rapid growth, therefore, income must be concentrated in the hands of the rich, whose marginal propensity to save is relatively high. Kaldor (1978) forms his explanation on the basis of the concepts of marginal propensity to save and consume. Poor people have a higher marginal propensity to consume which results in less investment and more consumption of goods. While investment aids economic growth by increasing the potential output through increases in efficiency and productivity consumption only has short-run effects. Kuznets (1955) explains the relationship between growth and inequality by highlighting the shift in labor from sectors with low productivity to sectors with high productivity which initially increases aggregate inequality substantially and only later decreases. The Kuznets Curve demonstrates an inverted U-shaped curve suggesting that inequality first increases and then decreases with development.\

However, how economists interpret the inverse relationship also differs. While Adelman and Morris (1973) argue that this pattern reflects a process of absolute impoverishment of lower-income groups in developing countries, Ahluwalia (1976) refers to a relatively more positive view that the worsening in relative inequality occurs not because there is a decline in the absolute incomes for the lower-income groups but because rates of growth of income for lower-income groups are lower than the rates for upper-income groups. These two opposing outlooks on this relationship occur because of the different approaches adopted by different economists during the process of analysis. Ravallion (2001) argues that given existing inequality, the income gains to the rich from distribution- neutral growth will be greater than the gains to the poor. Since wealth is more concentrated in the hands of the rich it is rational to assume that the rich will benefit more from growth compared to the poor. However, Ravallion (2001) states that the poor will still gain from growth. Moreover, empirical studies have shown that high growth rates don't necessarily cause any reduction in poverty and income distribution. Fields (2000) basing his conclusion on empirical studies, concludes that there is not any systematic (negative or positive) relationship between growth and inequality and it depends on the initial conditions of countries.\

Correlating with the empirical data of Dollar and Kray (2002) and Benabou (1996) it could be stated that on average, the rich will tend to capture a much larger share of the rise in national income from growth than the poor and income inequality and growth are inversely related. Voitchovsky (2005) argues that inequality at the top end of the distribution is positively associated with growth, while inequality lower down the distribution is negatively related to subsequent growth. She then emphasizes the misleading characteristic of single summary statistics, such as the Gini coefficient, which could reflect an average of the two offsetting effects. Similar to GDP and GNI, the Gini coefficient can be misleading because as a single indicator it aims to explore economic development and increases in the standard of living which is multidimensional. Therefore, while assessing the relationship between inequality and growth it is crucial to not only use GDP, Gini Coefficient but also use the indicators of health, education, etc.\

However, in the analysis of Barro (2000), inequality appears to encourage growth only within rich countries and to slow it down in poorer countries. This suggests that in countries where most of the population is living below the poverty line inequality does not function as an engine for growth. It has detrimental effects on the population because they are already living below the poverty lines determined by the World Bank. Moreover, the income generated by economic growth in poor countries is used up because of the high marginal propensity to consume. Poor people are more inclined to consume their income rather than invest in capital goods etc. Similarly, Mckay (1997) argues that if a country has a high inequality, growth might be less effective in reducing poverty. Bruno, Ravallion, and Squire (1998) confirm this argument analyzing the data of 17 countries showing that when holding the dispersion of income the same, higher income growth causes larger poverty reduction and holding the growth rate constant, as distribution becomes more dispersed, poverty reduces in smaller rates.\

On the other hand, some economists argue that economic growth and reduced inequality can go hand in hand. Birdsall, Ross, and Sabot (1995) explore the achievement of East Asian countries in both achieving high rates of economic growth and reduced inequality which is contrary to the conventional opinion that there is a tradeoff between inequality and growth.  While they highlight the importance of high levels of investment in education, they underline that the fact that East Asian development strategy promoted a dynamic agricultural sector and a labor-demanding, export-oriented growth path which also stimulated growth and reduced inequality. As Birdsall, Ross and Sabot (1995) highlight, it is important to reinforce the positive effect of rising education on growth by macroeconomic and trade policies, including an emphasis on manufactured exports, that generated demand for skilled labor. This illustrates the importance of investment in human capital along with physical capital. Investments in health care and education increase the skills and productivity of workers resulting in more efficient production, less unemployment and less inequality in the distribution of income.\

Turkey has one of the highest Gini coefficients (0.42) within the OECD countries. Turkey's inequality indicators such as the number of poor living at 3.20 dollars or 5.50 dollars a day fluctuate between years. Moreover, there has been limited research on the relationship between inequality and growth in Turkey. Bayar and Günçavdı (2016) attribute this occurrence to the lack of a period of uninterrupted economic growth in Turkish history and the availability of data for a substantially long time period. Research on this relationship has increased after the Turkish Statistic Institution (TurkStat) has made available relatively more reliable data on the budgetary records of Turkish households. This data is now available for the period between 2002 and 2009.\

Bayar and Günçavdı (2016) argue that economic growth has often been an economic priority in the Turkish politics, and it was believed that increased income in a growing economy, de facto, brought about an improvement in inequality as if it was a side effect of high economic growth policy. They claim that Turkey had preassumed that increases in the total value of output would also improve the distribution of income. The paper which assesses the distributional impact of macroeconomic policies under the AKP, reached the conclusion that the macroeconomic policies which bring high economic growth rates, lower inflation, and lower interest rates are a pre-condition for generating further improvement in inequality.  This suggests that there is not a consensus among economists on the relationship between inequality and growth. While Kuznets (1955) and Kaldor (1978) argue that economic growth worsens inequality initially Bayar and Günçavdı (2016) and Birdsall, Ross and Sabot (1995) demonstrate that some countries can achieve high growth rates and economic equality. Torul and Öztunalı (2018) focus on Turkey's income and wealth distribution in 2014. They highlight that while income inequality in Turkey has been stagnant over the last decade, recent evidence signals for an ever-increasing wealth concentration in Turkey, which reached alarming levels by 2014. This demonstrates a similar trend with the findings of Dollar and Kray (2002).  Moreover, Ucal, Bilgin, and Haug (2014) explored how foreign direct investment (FDI) and other determinants impacted income inequality in Turkey in the short- and long-run using the Auto-Regressive Distributed Lag (ARDL) modeling approach. They concluded that increasing FDI inflows have caused income inequality in Turkey to increase in the short run but not in the long run.  They also stated that an increase in the literacy and GDP growth grates reduces inequality both in the short and long run. This study implies that policies that place GDP growth alone at the center of reducing income inequality will be insufficient in the long run. Improving literacy rates (education) is crucial for a sustainable solution to income inequality, in addition to sustained economic growth. Duman (2008) also examines the link between educational variables and income inequality in Turkey. He argues that the dispersal of education among the income groups is rather high and in recent years there has been a growing gap between the educational expenditures of rich and poor households. He then elaborates on the government's expenditure on education is diminishing and becoming more biased towards tertiary education, which in turn decreases the chances of poor households utilizing these services. This suggests that education and how it is dispersed between the socio-economic groups have a significant effect on inequality. To achieve reduced income inequality Turkey should increase its investment spending in human capital, especially its investment in education. \

In addition to education, spending on social protection has tremendous effects on the distribution of income in Turkey. Tamkoc and Torul (2018) show that Turkey's wage, income and consumption inequalities all exhibit downward time-trends over the 2002-2016 period, which accord well with the rapid minimum wage and social protection spending growth over the period of interest. They demonstrate an increase from 9.33 \% in 2002 to 12.83 \% in 2016 in the share of GDP devoted to social protection expenditure. They also state that and increase in social protection expenditure also contributed to the downward time-trend in Turkey's income and consumption inequalities. Furthermore, Bakis and Polat (2013) highlight the importance of skill-biased technical change or minimum wage changes. Their findings suggest that during the period between 2002 and 2004, the relative supply of more educated workers to less-educated workers stayed almost constant while their relative wages have decreased in the benefit of less-educated workers. However, in the second period between 2004 and 2010 the relative supply of more educated workers to less-educated workers had risen while their relative wages remained constant or kept increasing in the benefit of more educated workers. Their results demonstrate the real minimum wage hike in 2004 corresponds to a major institutional change that proves to be welfare increasing in terms of wage inequality. It could be concluded that in addition to education, institutional changes through government intervention is able to steer the distribution of income towards a fairer path. \

\clearpage

\chapter{THE DATA}
\vspace*{0.4 cm}
\doublespacing
\noindent Gini index demonstrates the extent of income inequality in a country. The index is derived by measuring the area between the Lorenz curve and the perfect equality line. The Gini index is between 0, illustrating perfect income equality and 1, illustrating perfect income inequality. The index fluctuates between these two values. The annual Gini Index data from the years 2002-2017 are obtained from the website of the World Bank. World Bank states that the index is based on "primary household survey data obtained from government statistical agencies and World Bank country departments".  There are some limitations to the data. For instance, the data are not strictly comparable across countries or years due to the difference in surveys that make up the data and the definition of income in those surveys. Moreover, the difference in household size and the extent of income sharing among members can alter the Gini coefficient. However, World Bank aims to make the data as comparable as possible. \

Income share held by the highest 20 \% is the percentage share of income that is collected in the hands of each decile or quintile in a country. Perfect income equality requires each quintile to receive 20 \% of the real GDP. And the deviations between the perfect income equality shares illustrates the income inequality in a country. The annual data from the years 2002-2016 is obtained from the website of the World Bank. Similar to the Gini index, the data are obtained from the primary household survey carried out by government statistical agencies and World Bank country departments. Some limitations include lack of sufficient and high-quality data from low-income countries which results in low frequency and lack of comparability which then causes uncertainty over the magnitude of poverty reduction.\

GDP per capita is obtained by dividing the RGDP by the population. This value indicates the standards of living in a country. GDP per capita is the annual percentage growth rate of GDP per capita based on constant local currency. Positive growth of GDP per capita suggests that the standards of living in a country are increasing in accordance with the increase in RGDP per capita. The annual GDP per capita growth data from the years 2002-2017 are acquired from the website of the World Bank. The data is constructed from both the World Bank national accounts data, and OECD National Accounts data files. One limitation of GDP per capita growth is that it is difficult to do a cross country analysis. GDP per capita disregards the prices in the country which makes it hard to compare different countries with different price levels.\ 

The adult literacy rate is defined as the percentage of people ages 15 and above who can both read and write with understanding a short simple statement about their everyday life by the World Bank. The percentage of literate people sheds light on the economic development of a country. It is an indicator that helps to predict the quality of the future labor force. Moreover, high levels of literate females indirectly result in healthier and more educated children. The annual data between the years of 2004-2016, except 2008, is obtained from the website of the World Bank. However, the source of the data is the UNESCO Institute for Statistics. Limitations include difficulties in the cross-country analysis due to the differing definitions and methods of data collection in different countries.\

Net primary school enrolment is the percentage of children of official school age who are enrolled in a school. Primary education is crucial because it provides children with basic reading, writing and mathematics skills and an introductory level of understanding of various subjects. Therefore, it is also an indicator of the economic development of a country. The annual data between the years of 2004-2016, with the exception of 2008, is obtained from the website of the World Bank. However, the source of the data is the UNESCO Institute for Statistics. The source of the net enrollment rate is annual school surveys that do not reflect the actual attendance or dropout rates during the year. Therefore, is one of the limitations of the data in addition to the differing education lengths.\

Spending on education at early childhood and tertiary education levels as a percentage of GDP illustrates the total spending as a percentage of GDP by private and public institutions on these sub-groups of education. It covers expenditure on schools, universities and other public and private educational institutions. The data suggest the importance given on investment in human capital in a country, therefore, it can reflect the economic development of the country. The data are obtained from the website of the OECD. The data was only gathered between the years of 2014-2016 annually and the paper only took into consideration the percentage of GDP value in 2015 for all of the countries. One main limitation to the data is that it only covers the years 2014,2015 and 2016 which is not a long enough period to indicate the level of economic development in a country. \

Spending on education can be done by the government or/and private institutions. Government spending on education includes direct provision of educational institutions and public subsidies given to households to improve educational attainment. Public expenditure on education illustrates the priority given by governments to education compared to other areas of investment, such as health care, social security, defense, and security. High public investments in education increase income equality by increasing the number of well-educated individuals. Moreover, private spending on education is educational investments undertaken by private sources such as private institutions and households. The annual data on private and public spending from primary to tertiary education as a percentage of GDP is also obtained from the official website of the OECD. The data only included the time period of 2014-2016. This paper used only data from 2015 to obtain consistent data from each country. This data suffers from the same limitation as the total spending on education data. Besides, private spending does not include expenditure outside educational institutions such as private tutoring for students.\

Spending on health illustrates the spending on the final consumption of health care goods and services. Health care is financed by the government through government spending and compulsory health insurance and by private institutions, NGOs and households which are regarded as voluntary. The spending on health increase income equality by improving the efficiency and productivity of individuals. The annual data of total, voluntary and compulsory spending on health which is measured as a share of GDP is obtained from the official website of the OECD. The original data ranges between 2014-2018 in countries and the spending in 2018 was analyzed in this paper. The limited range of data and the difference between the latest data available increase the difficulty in both measuring development in a country and contrasting the values across countries.\
 
\newpage
\chapter{GDP GROWTH and INCOME EQUALITY}
\vspace*{0.4 cm}
\doublespacing
\noindent After the economic crisis in 2001, Turkey implemented a structural adjustment program to overcome the economic problems caused by the crises under the supervision and technical support of the International Monetary Fund and the World Bank (Albayrak, 2009). Bakis and Polat (2013) illustrate that following the launch of the IMF recovery program, economic growth was smoother and higher on average, the ratio of public debt to GDP declined substantially, and inflation and real interest rate decreased drastically. Taylor (2004) concludes that the fiscal discipline involved in these types of programs aims to attain primary fiscal surpluses which in turn result in two major problems. Firstly, fiscal policies affect the components of aggregate demand of an economy. They can directly impact government spending, or they can increase investment and consumption components of aggregate demand by influencing taxes. The direct result of the IMF determined fiscal policy is a reduction in public spending including expenditure on education and health. This widens the gap within different income groups because while some are dependent on public health and education, some are able to afford private institutions. Moreover, the changes in the composition of tax structure also create direct and indirect redistributive effects on different household groups via price and income changes (Albayrak, 2009). For instance, indirect taxes such as excise taxes are regressive which suggests that the average tax rate decreases as the amount subject to taxation increases. This increases the divergence between low and high-income earners further causing income inequality. Furthermore, these policies that were adopted resulted in rapid growth rates (8 \%) in the Turkish economy in 2002. However, despite these high growth rates the economic crisis in 2001 the income inequality persisted.\

\begin{figure}[H]
	\centering
	\includegraphics[scale=0.4]{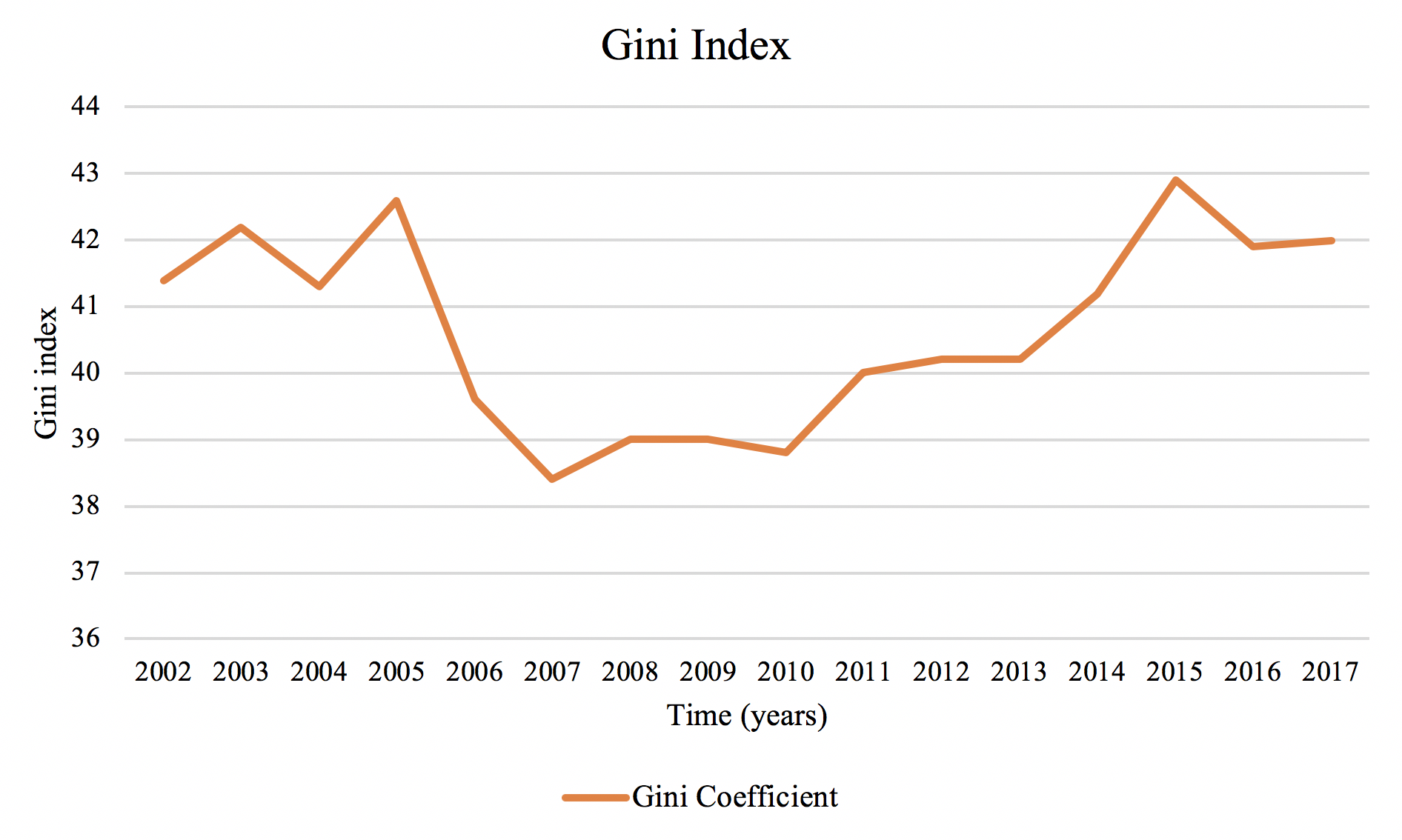}
	\caption{Gini Coefficient In Turkey}
\end{figure}

Figure 1 illustrates the changes in Turkey's Gini Coefficient between the years of 2002-2017. The graph has a U-shape suggesting that income inequality in Turkey first decreased then increased in the time period of 2002-2017. It can be concluded that Turkey's Gini coefficient experienced a rapid decrease between the years of 2005-2007. This suggests that Turkey's income distribution improved over that time period. Bayar and Günçavdı (2016) relate this decrease in the Gini coefficient with the income generation impact of AKP economic policies during the full period. They assert that the policies became useful for labor income earner households, and the inequality within this group declined from 0.40 in 2002 to 0.37 in 2007. Similarly, Tamkoc and Torul (2018) illustrate that the downward trend in the Gini coefficient accord well with the rapid minimum wage and social protection spending growth over the period of interest. They then argue that the 119 \% growth in net minimum real wage has served to mitigate Turkey's economic inequalities. Moreover, the increase in the share of GDP devoted to social protection expenditure which increased from 9.33 \% in 2002 to 12.83 \% in 2016 is a major factor that affects income distribution. Governments undertake social protection expenditures to establish income equality between their citizens. The policies redistribute income by collecting greater tax revenue from high-income earners and providing merit goods to the public.  Education and health care which are fundamental basic rights can, therefore, be accessible to all of the population. Still, Turkey's lowest Gini coefficient which was 0.384 in 2007 is relatively high compared to the other member countries of the OECD.\

This rapid decrease was followed by a steady increase in the Gini Coefficient between the years of 2008-2015. Similar to the crisis in 2001, after the Great Recession in 2008, the Turkish economy slowed down at first and shrank radically after (by growing 0.85 \% in 2008 and -4.70 \% in 2009), which coincides with an upward trend in the Gini coefficient (Tamkoc and Torul, 2018). This illustrates that income equality in Turkey worsened over that duration of time. Moreover, the share of debt interest payments increased in time and it was around 50 \% for the 2000s. As these expenditures go to higher-income households who have savings to lend to the government by buying government bonds, they are expected to have a widening impact on the distribution of income (Albayrak, 2009).\
 
\begin{figure}[H]
	\centering
	\includegraphics[scale=0.4]{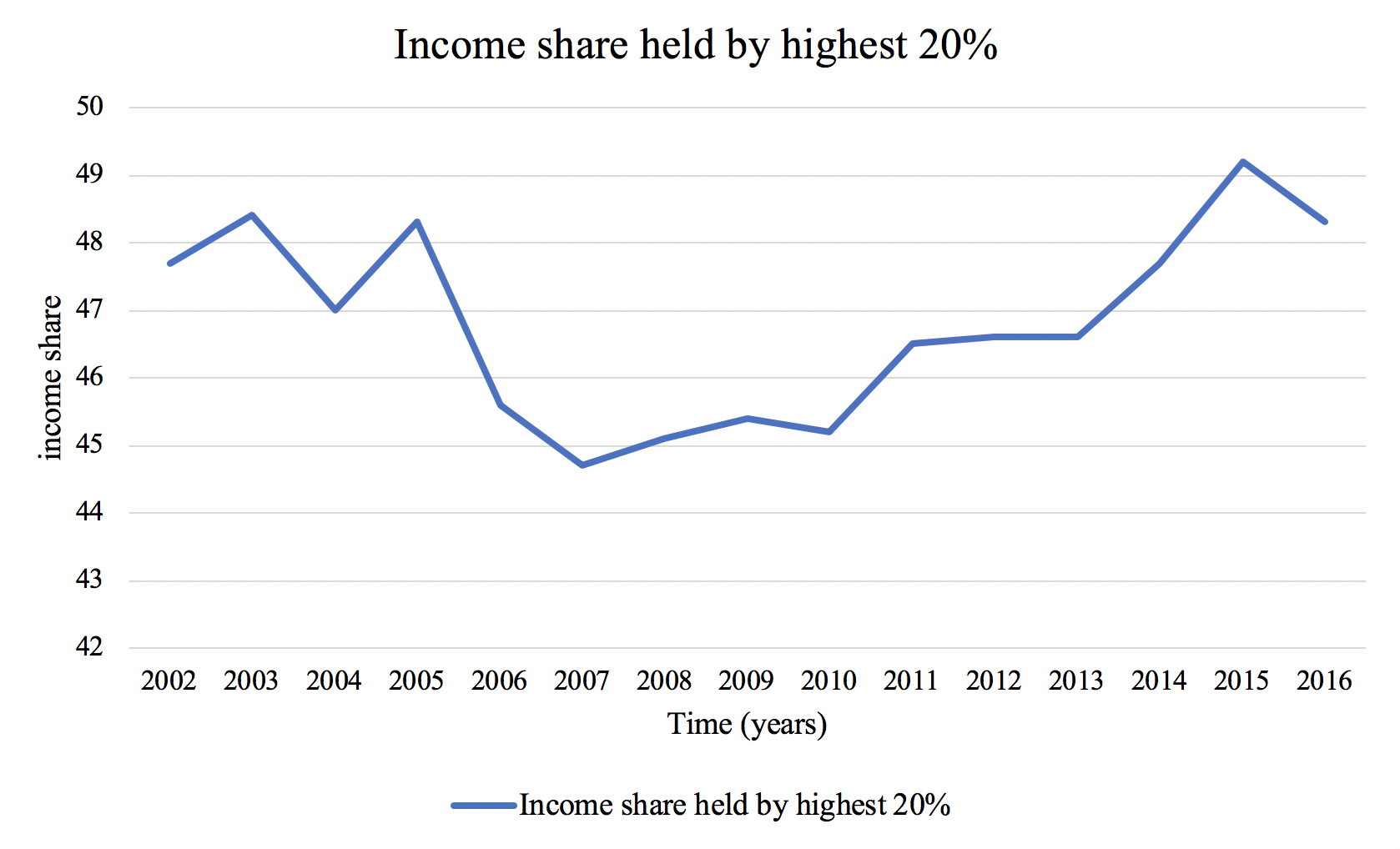}
	\caption{Income share held by the highest 20 \% in Turkey}
\end{figure}

Figure 2 demonstrates the income share held by the highest 20 \% in Turkey. Between the time periods of 2002 and 2016 income share held by Turkey's highest 20 \% fluctuates around between 45 \% and 50 \%. However, the income share held by the highest 20 \% starts to decrease in 2005 and experiences its lowest value in 2007 (44.7 \%). Then, it demonstrates an increasing trend from 2008 to 2015. It experiences its highest value in 2015 (49.2 \%). Nearly half of the income in Turkey was held by the highest 20 \% between 2002-2016. This suggests that there is a highly unequal distribution income in Turkey. Moreover, this graph demonstrates a similar U-shape with the Gini coefficient. Gini coefficient and the income share held by the highest 20 \% correlate because as the income share held by the highest 20 \% increases the Gini Coefficient increases therefore, income equality worsens. \
 
 \begin{figure}[H]
	\centering
	\includegraphics[scale=0.4]{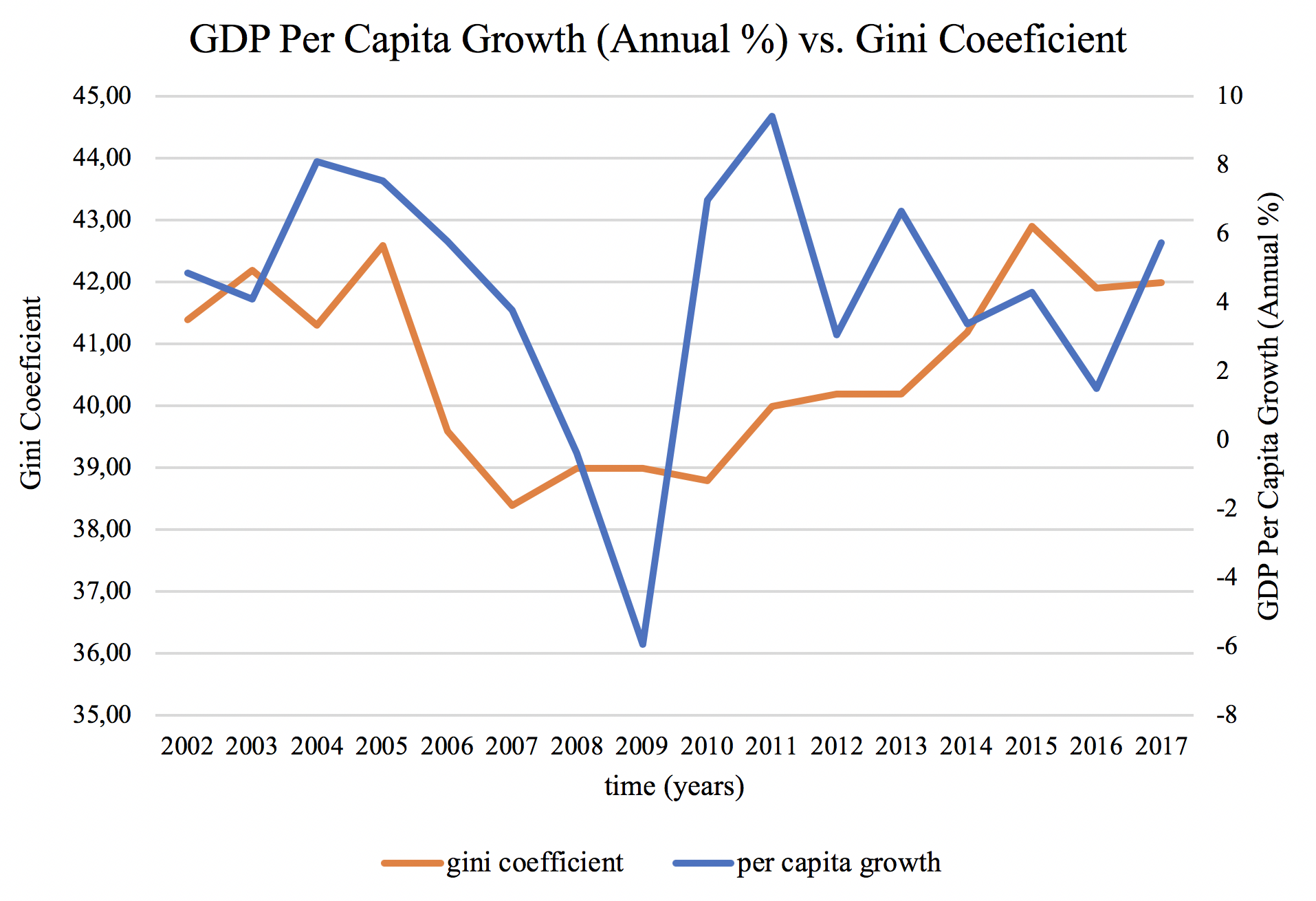}
	\caption{GDP per capita growth vs. Gini Coefficient in Turkey}
\end{figure}

The graph above illustrates the relationship between the GDP per capita growth and the Gini Coefficient in Turkey. It demonstrates a positive relationship between GDP per capita growth and the Gini coefficient. The lowest GDP per capita growth is accompanied by the lowest Gini Coefficient (2009). This demonstrates an inverse relationship between economic growth and inequality. However, as it was discussed above GDP per capita is not the only factor that affects income equality. For example, between 2004 and 2005 income equality worsens and GDP per capita growth decreases. The composition of growth is also crucial. The growth of sectors that result in a higher GDP per capita can widen or shrink the gap between high and low-income earners. Therefore, even though the graph demonstrates mostly a positive relationship between economic growth and the Gini coefficient, it is impossible to conclude that the increase in GDP per capita growth is a driving factor of inequality. There is no consensus on the effects of GDP per capita growth on income equality. The other major reason is that if GDP per capita growth increases due to the increases in social expenditure on health care and education the increase in GDP per capita growth might work in favor of income equality. \

\clearpage

\chapter{EDUCATION SPENDING and INCOME INEQUALITY}
\vspace*{0.4 cm}
\doublespacing
\noindent Education and health services are important in terms of their positive impact on growth and development as well as their redistributive power, and it has been shown that public expenditures on these services' narrow inequalities significantly (OECD, 2008: Chapter 9). Well-established education and health care systems result in more efficient, productive and skilled labor. Moreover, the recent empirical literature for developing countries focuses on public spending programs that have the explicit goal of improving distributional equity and confer a personal benefit upon users, such as education, health, infrastructure services and social transfers (Albayrak, 2009). \

In Turkey education is compulsory for children between the ages of 6-14 so that children from all income groups can afford basic education, education is provided by the government free of charge. However, there are private schools in Turkey that are offered by private institutions that charge a high price for their services. The difference between the quality of private and public education in Turkey is a major factor that contributes to income inequality in Turkey. The children who go to private schools not only have more qualified teachers but also, they learn a second language such as English from a young age. This results in more skilled students graduating from private schools compared to public schools. Moreover, there is also a difference in the number of students between high- and low-income earners in universities. The children who go to private schools are taught in classes that have almost half the number of students in public schools. While the average maximum number of students in a private school classroom is 20 the number doubles in public schools. Naturally, since there are more children in the class there is less attention given to each child about their progress. This suggests a flaw in the education spending of Turkey. More emphasis is put on tertiary education compared to primary and early childhood education. This results in a lack of strong foundation for students.\

\begin{figure}[H]
	\centering
	\includegraphics[scale=0.4]{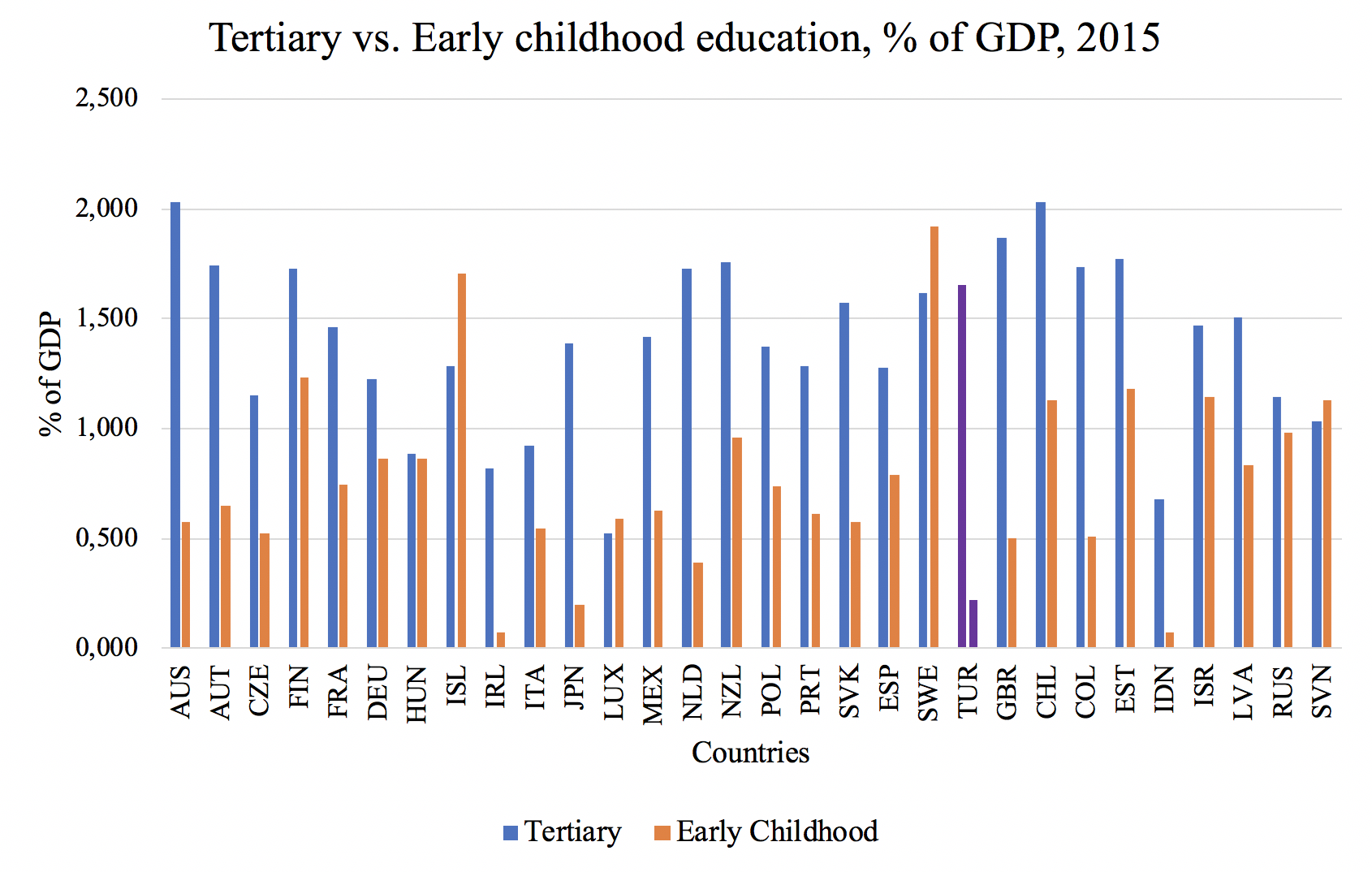}
	\caption{Spending on education, tertiary and early childhood education, \% of GDP, 2015}
\end{figure}

Figure 4 illustrates the \% of GDP that is spent on education in many countries in 2015. The figure illustrates that Turkey has spent 0.22 \% of its GDP on early childhood education and 1.65 \% of its GDP on tertiary education. While Turkey has relatively high spending on tertiary education; it has very low spending on early childhood. Together with Mexico, the United States, and Canada, Turkey exhibits the highest college premium in the early 2000s (Tamkoc and Torul, 2018). While Turkey's spending on tertiary education demonstrates similar values with Denmark (1.69), Sweden (1.62) and Finland (1.73) Turkey's spending on early childhood education demonstrates contrary values with these three countries Denmark (1.26), Sweden (1.92) and Finland (1.24). Duman (2008) states that "private and social returns to primary and secondary schooling turn out to be quite high in Turkey therefore, more spending in these areas could improve the education and earning disparities. However, government expenditure on education is diminishing and becoming more biased towards tertiary education, which in turn decreases the chances of poor households utilizing these services." Therefore, it raises the question of whether Turkey invests in the right level of education. \

\begin{figure}[H]
	\centering
	\includegraphics[scale=0.4]{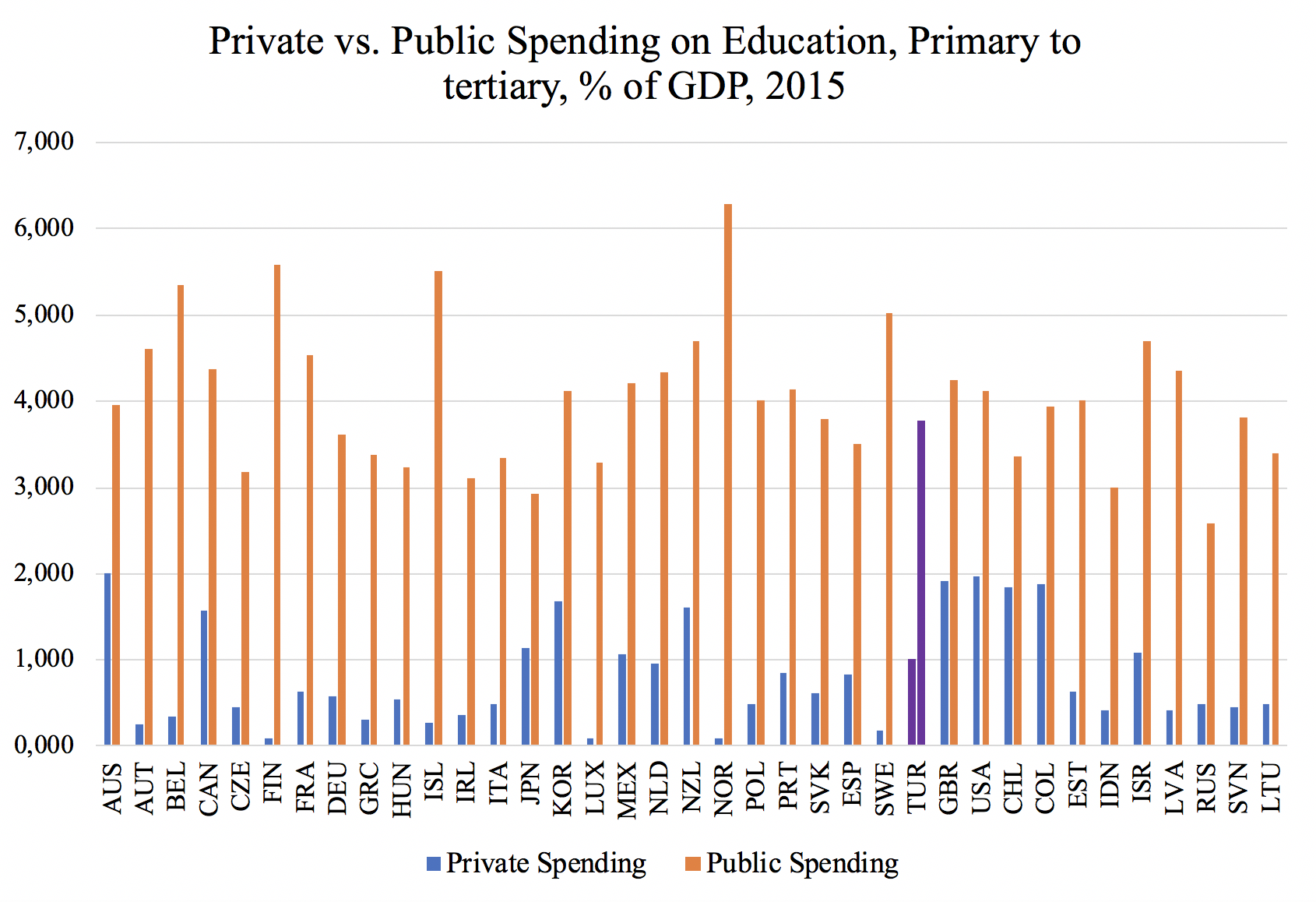}
	\caption{Private vs. Public Spending on Education, Primary to Tertiary, \% of GDP, 2015}
\end{figure}

Figure 5 demonstrates Turkey's private and public investment in education from primary to tertiary level in 2015. As can be seen from figure 2, Turkey has high private spending on primary to tertiary education. On the other hand, Turkey has low public spending on primary to tertiary education. The difference between public and private spending is the major reason for inequality between socio-economic classes in Turkey. The distributional effects of public education services in Turkey were investigated by Pinar (2004) for 1994 and 2002. The paper found that primary and secondary education is pro-poor, but this pro-poor impact declines when private expenditures on education are included in the analysis. The quality of private and public education has distinct differences and low-income groups are not able to afford private schooling due to high prices. This worsens inequality because it widens the gap between high income and low-income groups. Duman (2008) supports this view by arguing that due to the limited public spending on primary and secondary education and growing private spending, the spread between socio-economic groups will not be decreased significantly. Moreover, in addition to private schooling private tutoring carries great importance because of its positive effects on university placements. Tansel and Bircan (2008) examines the determinants of receiving private tutoring and getting placed in a university program by using the 2002 survey of the applicants to the university entrance examination and their performance in OSS. The findings of the paper show that parents' education level, employment status, and income level affect positively the possibility to receive private tutoring and also to get placed in a university program. Since private tutoring is most available for high-income earners it further increases the divergence between low and high-income earners. Although public education services dominate the education system in Turkey, out-of-pocket expenditures on education may still create barriers to utilizing public education services for people with low-income groups, which we have seen clearly for higher education services above. Even if the tuition fees for public universities are nominal in Turkey, getting placed in a public university requires household private expenditures on education (Albayrak, 2009). \

\begin{figure}[H]
	\centering
	\includegraphics[scale=0.4]{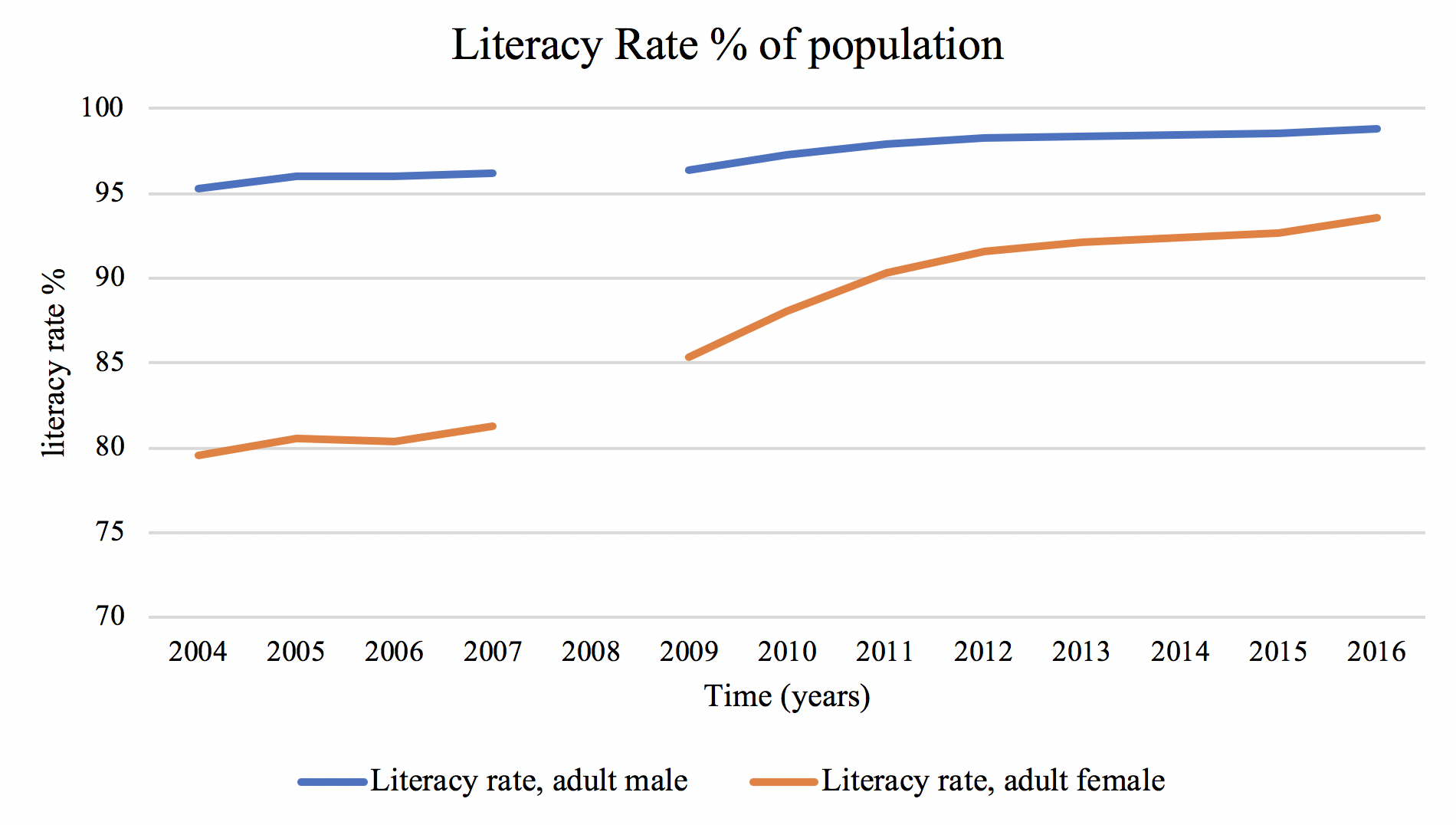}
	\caption{Literacy rate \% of the population, adult male and adult female in Turkey}
\end{figure}

Moreover, Figure 6 demonstrates the adult literacy rate \% of males and females in Turkey. The percentage of the male population who is literate illustrates a steady increase. On the other hand, the percentage of the female population who is literate experiences a steeper growth. However, there is still a difference between the percentage of females and males that are literate. In 2016 the gap was more than 5 \%. But the graph demonstrates that in the early 2000s there was a greater gender gap in literacy rate. In 2004 the difference was greater than 15 \% which suggests that Turkey not only faced income inequality but also gender inequality. Moreover, whether an individual is literate reflects on their employability. The income share of workers with a low level of education in total incomes has reduced overtime in Turkey, whereas earnings of people with a university degree have significantly increased in recent years (Albayrak, 2009).  It has been shown that education has positive impacts on labor force participation and earnings in Turkey. Education increases labor force participation, and this effect and the returns to education increase with the level of education in Turkey. (Albayrak, 2009) Therefore, in the long run, it creates a divergence between the incomes of males and females. However, if the next several years follow the same trend this gap will eventually close and the literacy rate will increase beyond 99 \%.\
 
\begin{figure}[H]
	\centering
	\includegraphics[scale=0.4]{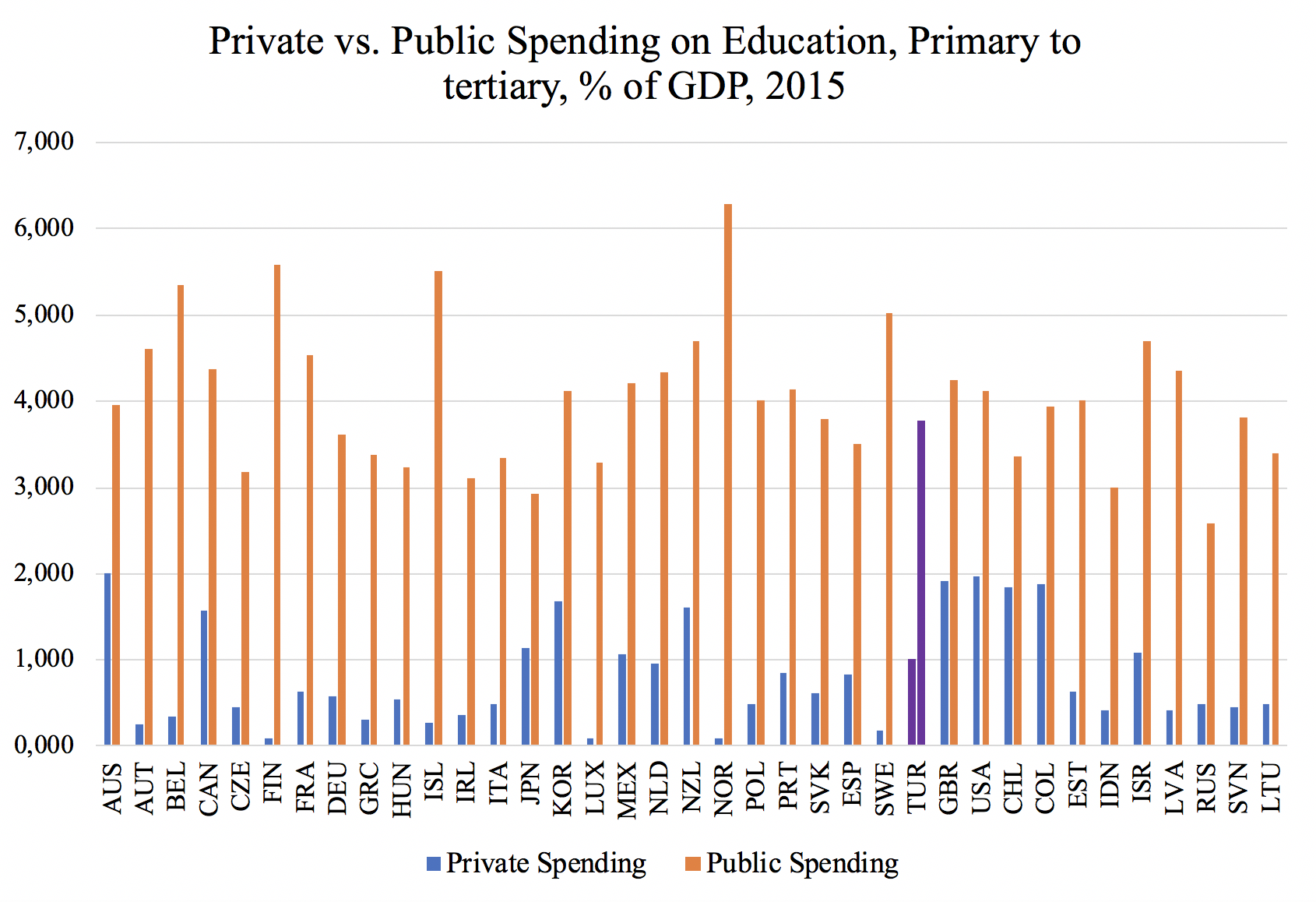}
	\caption{Primary school enrollment \% of the population, male and female in Turkey}
\end{figure}

The figure above demonstrates the \% of net primary school enrollment of females and males in Turkey. There is a constant difference rate between male and female enrollment in primary education during the time period of 2006-2016. While 88 \% of female primary-school-age children in the first quintile go to school, this rate is 95 \% for male primary-school-age children (Albayrak, 2009). One of the reasons why female enrollment is lower in Turkey is due to Turkish customs. Low-income families in Turkey prioritize the education of their male children above their female children. In order to increase the female enrollment in primary education Doğan Holding started a campaign called "Baba Beni Okula Gönder" in 2005. This campaign which was undertaken by a private firm aimed to raise awareness and provide scholarships to the female children in Turkey. Moreover, this difference is not only present in primary education but tertiary education as well. Albayrak (2009) demonstrates that the female enrollment rate for higher education is 4.5 \% for the poorest quintile, whereas the female enrollment rate for higher education is 49 \% for the richest quintile. Overall, the primary school enrollment rate differences between males and females could be another factor of income inequality in Turkey.\

\clearpage

\chapter{HEALTHCARE SPENDING and INCOME INEQUALITY}
\vspace*{0.4 cm}
\doublespacing
\noindent The Ministry of Health is the main government body responsible for health sector policymaking, implementation of the national health strategies and the largest health services provider in Turkey. Similar to education, Turkey has both public and private health facilities. Unfortunately, there is a difference between the quality of health care received between the institutions. Public institutions are mostly overcrowded which causes the patients not to get the service they need.\

\begin{figure}[H]
	\centering
	\includegraphics[scale=0.4]{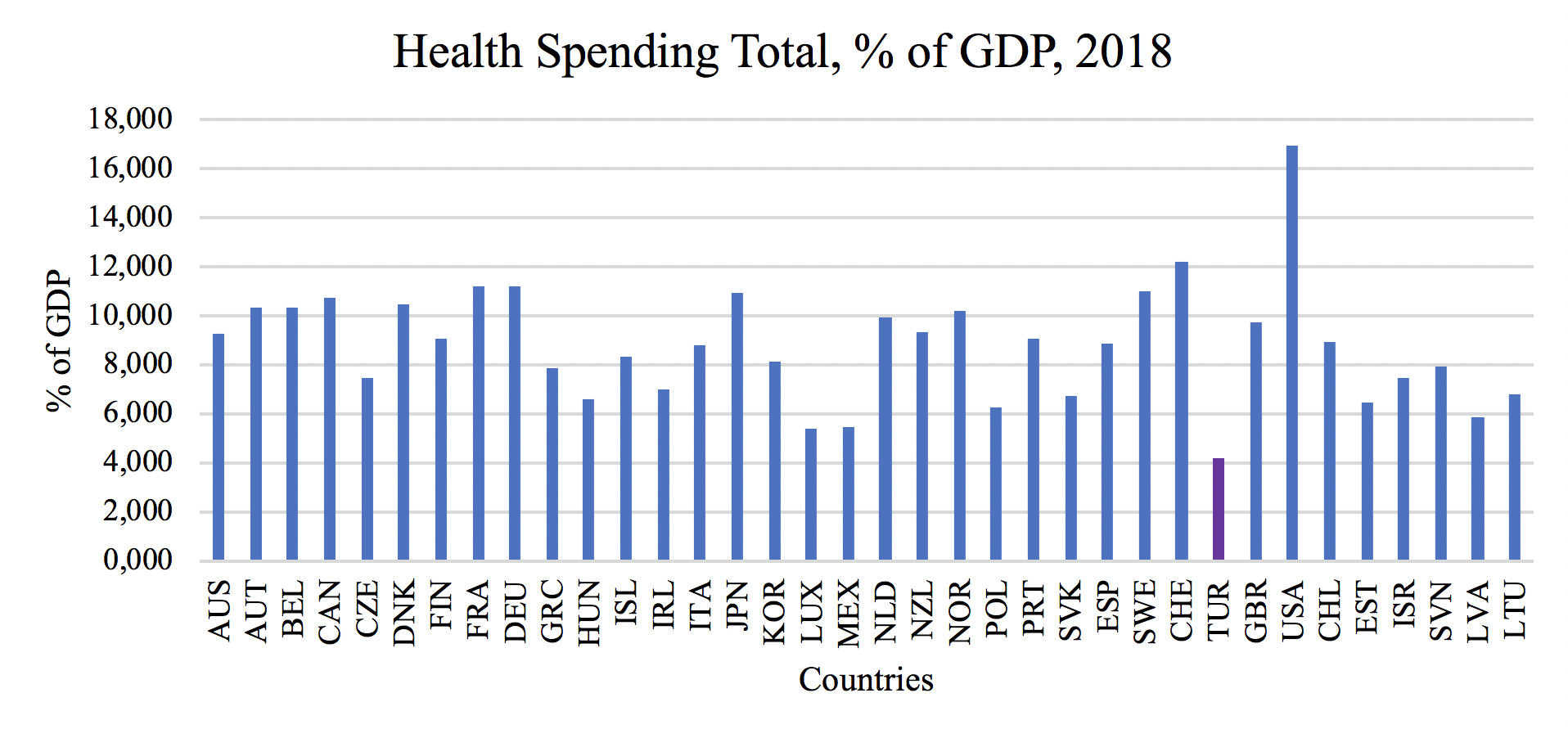}
	\caption{Health Spending Total, \% of GDP, 2018 }
\end{figure}

TUIK disclosed Turkey's expenditure on health care in 2016 and 2017. Turkey allocated 4.6 \% of its GDP on healthcare in 2016 and this number decreased to 4.5 \% in the following year. This highlights the three main problems of the Turkish public health system which are the absence of universal coverage for health services, inadequate public spending on health services and low quality of services. Figure 8 demonstrates the data from the OECD of Turkey's total spending on health care in 2018 which supports the second main problem of the health care system in Turkey. With 4.2 \% of its GDP spent on health care, Turkey is the third country to least invest in health care. This suggests that Turkey's investments in health care have been decreasing since 2016. Moreover, almost 78 \% of the households who spent money on health care have at least one individual in the family with health insurance, implying the importance of out-of-pocket expenditures in the health system (Albayrak, 2009). According to MoH (2006), 16 \% of people with a health problem did not do anything, for 60 \% of these it was due to lack of money, and most were living in eastern regions, rural areas and uninsured. This demonstrates the lack of investment undertaken by the government on health care. Moreover, the need for out-of-pocket expenditures suggests that health care might not be accessible to everyone. This increases the divergence between the poor and the rich because the rich are able to access better health care and achieve higher rates of productivity. Furthermore, drugs are the highest expenditure component followed by doctors and inpatient care in Turkey. Some drugs including some cancer medication are not covered by insurance. This suggests that low-income earners are not able to afford medicines that are vital for their well-being. Without the investment and direct provision of the government, these patients are unable to afford these products which could result in life-threating problems. Investments in health care similar to education are investments in human capital. Better health conditions increase workers' morale, productivity, and efficiency. Therefore, the low spending of expenditure on the healthcare system is one of the major factors of high-income inequality in Turkey.\

\begin{figure}[H]
	\centering
	\includegraphics[scale=0.4]{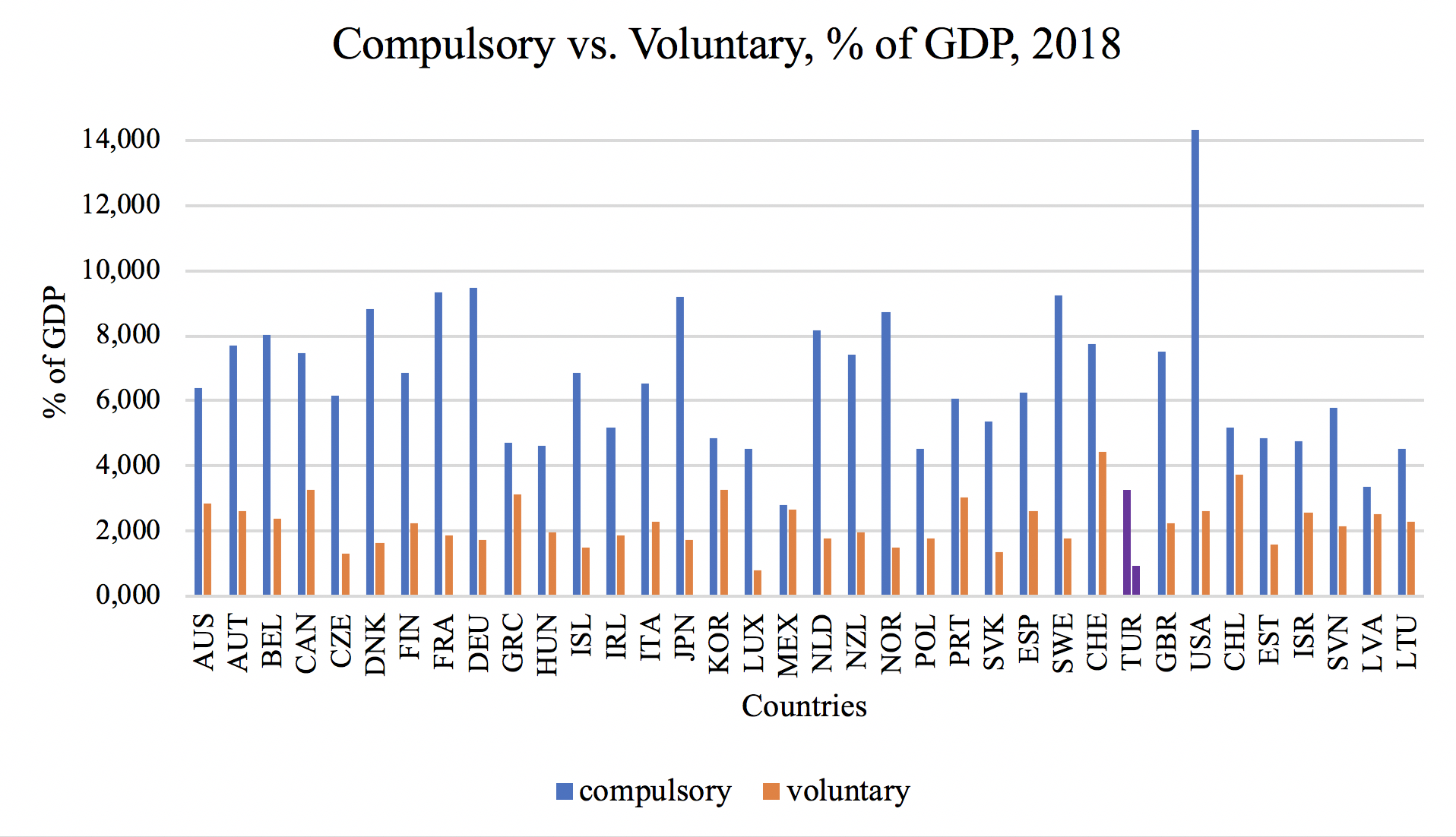}
	\caption{ Voluntary and Government Spending on Health Care, \% of GDP, 2018 }
\end{figure}

Figure 9 demonstrates the difference between voluntary and governmental expenditure on healthcare. Even though Turkey is the second to last country to voluntarily invest in health care following Luxemburg (0.8), its expenditure value is not very far off from other countries compared to its government expenditure. While Turkey's government expenditure on health care is 3.3 \% of its GDP Luxemburg's expenditure is 4.5 \% and many governments at least spent 4 \% of their GDP on health care. In 2018 OECD measured Turkey's GDP as 28 384 dollars which is similar to Greece (29 592 dollars) and Hungary (30 652 dollars). While government spending on health care in these countries is 4.7 and 4.6, respectively, Turkey's is much less. Therefore, it can be concluded that there is a huge gap in private and public healthcare spending in Turkey. The wide gap between private and public spending worsens inequality because the poor are unable to afford private health care services. This works to widen the gap between the poor and the rich.\
    
\pagebreak
\chapter{CONCLUSION}
\vspace*{0.4 cm}
\noindent Turkey has one of the highest Gini coefficients among the member OECD countries and it has experienced fluctuating Gini Coefficients between 2002 and 2017. This research deeply analyzed the sources of income inequality utilizing the comprehensive data from the World Bank, OECD and Turkish Statistical Institute.\

Firstly, the relationship between inequality and economic growth in Turkey is analyzed. The main observation is that until the crisis in 2008 there had been an improvement in the Gini coefficient. However, after the crisis, a downturn began hurting income equality in Turkey. The Gini coefficient started to increase and reach its peak in 2016. Even though it seems that there was a correlation between the Gini coefficient and the GDP per capita growth, it is shown that it is not possible to conclude that GDP per capita has a direct impact on income equality. Therefore, the effect of various factors in income inequality is analyzed.\

The results suggest that the literacy rate and enrollment in primary education contributed to the high levels of income inequality in Turkey. The biggest issue in Turkey is that education investments are directed to tertiary education rather than primary and early childhood education. However, if we look at economically developed countries, investment in education has a significantly higher \% rate of their GDP on primary and early childhood education. A high-quality primary education builds a strong foundation for further studies that develop over previous knowledge. Therefore, for the Turkish citizens to enjoy the advantages of investments in tertiary education they need to first benefit from primary education. However, low primary school enrollment in Turkey suggests that the government misallocates its resources and benefits that arise from education do not reach a significant amount of individuals to improve equality. This assertion is a valuable asset to Turkish policymakers in the field of education on topics such as where to spend government revenue to obtain the maximum benefit in terms of income equality and development. By reallocating the sources for education, the Turkish government could shrink the gap between high and low-income earners. Moreover, the undertaker of educational investments is also very important in terms of inequality. In Turkey, while public education is free of charge, the education offered by private institutions is very costly. When investments by private institutions increase and public investments stay stagnant, the difference between high and low-income earners increases further widening income inequality. Therefore, even though private investments are still valuable, the burden should be on the Turkish government to provide all of its citizens an adequate education for an economically more equal society.\

Moreover, even though Turkey's health care system is almost free of charge in most public institutions, Turkey is still one of the countries that invest the least of its GDP in healthcare. In addition, the ratio between public and private healthcare expenditure in Turkey is lower compared to other OECD members. Therefore, the low level of public expenditure and the considerable high level of private expenditure worsens the income inequality in Turkey. The argument is as follows: Private institutions charge a fee; therefore, they have more qualified staff and better equipment which results in better health care services. The paper highlights the implications of investment in health care in terms of income equality and concludes that the Turkish healthcare expenditure structure is one of the major reasons for income inequality.

\chapter*{REFERENCES}

\vspace{1cm}

 \singlespacing
\noindent Adelman I. \& C. T. Morris. (1973).). Economic Growth and Social Equity in \

Developing Countries,Stanford CA: Stanford University Press. 

 \singlespacing
\noindent Ahluwalia, M. (1976a). Income Distribution and Development. \textit{American Economic}\ 

\textit{Review}, 66(5), 128–135. \
 
 \singlespacing
\noindent Ahluwalia, M. (1976b). Inequality, Poverty and Development. \textit {Journal of} \ 

\textit {Development Economics},3, 307–342.

\singlespacing
\noindent Albayrak, O. (2009). The redistributive effects of fiscal policies in Turkey, 2003. PhD \ 

thesis, University of Nottingham, 2-15.

\singlespacing 
\noindent  Bakis, O. \& S. Polat. (2013). Wage Inequality in Turkey: 2002-2010. \textit{TUSIAD-Sabanci}\

\textit{University Competitiveness Forum}, 2, 1-5. 

\singlespacing 
\noindent Barro, R. J. (2000). Inequality and Growth in a Panel of Countries \textit{Journal of} \ 

\textit{Economic Growth},5, 5–32. 

\singlespacing 
\noindent Bayar, A. \& Günçavdı, O. (2016). Economic Reforms and Income Distribution in \ 

Turkey. 1-28. 

\singlespacing
\noindent Be ́nabou, R. (1996). Inequality and Growth.  \textit{NBER Macroeconomics Annual}\

11, 11–73.

\singlespacing
\noindent Birdsall, N., Ross, D., \& Sabot, R. (1995). Inequality and growth reconsidered: lessons \ 

from East Asia.\textit {World Bank Economic Review}, 9 (3), 477-508. 

\singlespacing
\noindent Bruno, M., M. Ravallion \& L. Squire (1999). Equity and Growth in Developing \ 

Countries: Old and New Perspectives on the Policy Issues.  \textit{World Bank Policy} \ 

\textit{Research Working Paper}(Report No. 1563 ). 

\singlespacing
\noindent Dollar, D., \& A. Kraay. (2002). Growth is Good for the Poor. \textit{Journal of Economic} \

\textit{Growth},7, 195–225. 

\singlespacing
\noindent Duman, A. (2008). Education and Income Inequality in Turkey: Does Schooling \ 

Matter?  \textit {Financial Theory and Practice}, 32 (3), 369 – 385.

\singlespacing
\noindent Fields, G.s. (2000). The Dynamics of Poverty, Inequality, and Economic Well-being: \ 

African Economic Growth in Comparative Perspective.  \textit {Journal of African Economies},9,\ 

45-78. 

\singlespacing
\noindent Kaldor, N. (1978). Capital Accumulation and Economic Growth. In Nicholas Kaldor, \ 

ed., Further Essays on Economic Theory. New York: Holmes and Meier \ 

Publishers, Inc.

\singlespacing
\noindent Kuznets, S. (1955). Economic Growth and Income Inequality. \textit{American Economic}\ 

\textit{Review},45, 1-28. 

\singlespacing
\noindent McKay, A. (1997). Poverty Reduction through Economic Growth: Some Issues,\

\textit{Journal of International Development}, 9, 665-673.

\singlespacing
\noindent Ravallion, M, (2001). Growth, Inequality and Poverty: Looking Beyond Averages, \ 

\textit {World Development},29 (11), 1803-1815. 

\singlespacing 
\noindent Tamkoc, N. \& Torul, O. (2018). Cross-Sectional Facts for Macroeconomists: Wage,\ 

Income and Consumption Inequality in Turkey.  \textit{Bogazici University, Working} \

\textit{Papers}, 2-20.

\singlespacing 
\noindent Taylor, L. (2004). Reconstructing Macroeconomics: Structuralist Proposals and \ 

Critiques of the Mainstream, Harvard University Press.

\singlespacing
\noindent Torul, O. \& Öztunalı, O. (2018). On Income and Wealth Inequality in Turkey. \textit{Central} \

\textit{Bank Review}, 18(3), 95 – 106. 

\singlespacing
\noindent Ucal, M., Bilgin M. H.\& Haug, A. A. (2014). Income Inequality and FDI: Evidence \ 

with Turkish Data. \textit{University of Otago Economics Discussion Papers}, No. 1407,\  

2-21.

\singlespacing
\noindent Voitchovsky, S. (2005). Does the Profile of Income Inequality Matter for Economic \ 

Growth?: Distinguishing Between the Effects of Inequality in Different Parts of

\ the Income Distribution, \textit{Journal of Economic Growth},10, 273–296.

\end{document}